\documentclass[journal]{IEEEtran}

\usepackage{amsmath,amsfonts}
\usepackage{algorithmic}
\usepackage{array}
\usepackage[caption=false,font=normalsize,labelfont=sf,textfont=sf]{subfig}
\usepackage{textcomp}
\usepackage{stfloats}
\usepackage{url}
\usepackage{verbatim}
\usepackage{graphicx}
\usepackage{framed}
\usepackage{multirow}
\usepackage{booktabs}
\usepackage{xcolor}
\usepackage{cite}
\usepackage{hyperref}

\AtBeginDocument{%
  }

\begin{document}

\title{Reinforcement Learning for Mutation Operator Selection in Automated Program Repair}
\author{Carol Hanna$^1$, Aymeric Blot$^2$, Justyna Petke$^1$\\
$^1$University College London, UK, $^2$Universit{\'e} de Rennes, France}


\maketitle

\begin{abstract}

Automated program repair techniques aim to aid software developers with the challenging task of fixing bugs. 
In heuristic-based program repair, a search space of program variants, created via mutations on software, is explored to find potential patches for bugs. Most commonly, every selection of a mutation operator during search is performed uniformly at random, whcih can generate many buggy, even uncompilable program variants. Our goal is to reduce the generation of variants that do not compile or break intended functionality which waste considerable resources. 
% and limits these approaches from being applied at scale.
In this paper, we investigate the feasibility of a reinforcement learning-based approach for the selection of mutation operators in heuristic-based program repair. Our proposed approach is programming language, granularity-level, and search strategy agnostic and allows for easy augmentation into existing heuristic-based repair tools.

We conduct an extensive empirical evaluation of four operator selection techniques, two reward types, two credit assignment strategies, two integration methods, and three sets of mutation operators using 30,080 independent repair attempts. We evaluate our approach on 353 real-world bugs from the Defects4J benchmark.\looseness=-1
The reinforcement learning-based mutation operator selection results in a higher number of test-passing variants, but does not exhibit a noticeable improvement in the number of bugs patched in comparison with the baseline, which uses random selection. While reinforcement learning has been previously shown to be successful in improving the search of evolutionary algorithms, often used in heuristic-based program repair, it has not shown such improvements when applied to this area of research.

% Our study shows that the epsilon-greedy multi-armed bandit algorithm with average credit assignment is best for mutation operator selection. 
\end{abstract}


\section{Introduction}
\label{intro}
Fixing bugs remains a largely manual and tedious process that often requires more time than is available to software developers~\cite{Wei2007,Bohme2017}. With the fast-evolving industry, this leads to the deployment of bug-prone products in an attempt to meet release deadlines~\cite{Liu2021,Tassey2002}. Research in the area of automated program repair (APR) aims to address this issue by automating the process of finding suitable patches for bugs in software systems. \looseness=-1
% However, despite the dire need for such tools, there hasn't been a wider adoption of APR techniques in industrial settings. While APR tools have seen a lot of success, their efficacy and efficiency rates remain too low for industry standards~\cite{Noda2020}.

The classification of APR approaches differs in the literature. According to a recent survey~\cite{surveyLeGoues}, the proposed approaches in the field of APR span across three main areas: constraint-based, learning-based, and heuristic-based techniques. Constraint-based approaches use the semantics of the buggy program to produce a constraint, then synthesise repairs that satisfy it~\cite{Xuan2018}.
Learning-based end-to-end repair techniques predict patches for buggy programs by learning features of the faulty code sections as well as their correct (developer-written) fixes~\cite{Chen2022}.
%Among APR approaches, heuristic-based APR has seen the earliest and thus far most industrial uptake:
%heuristic-based APR was first applied to software during the development of a management system for a medical application~\cite{Janus}. The earliest deployment for the purpose of bug fix automation at scale was Meta's SapFix~\cite{sapfix}, which combines various approaches including a single generation search over the space of higher order mutants. More recently, Bloomberg applied heuristic-based APR within their infrastructure to target simple and frequently occurring bugs~\cite{Kirbas2020}.
Among APR approaches, heuristic-based APR has seen the earliest and thus far most industrial uptake, including being first applied in the context of a management system for a medical application~\cite{Janus}, or more recently being used for automated end-to-end repair at scale at Meta~\cite{sapfix} or targeting frequently occurring bugs at Bloomberg~\cite{Kirbas2020}.

Heuristic-based APR uses search strategies, such as genetic programming~\cite{Koza1994} or local search~\cite{Hoos2004}, to navigate the space of software variants. These approaches require a test oracle for the buggy program to assess the correctness of the software variants. The oracle contains passing test cases and at least one failing test case that demonstrates the bug. As a first step, fault localisation techniques are employed to pinpoint the suspicious code sections. A program variant is created through the selection of a suspicious code location as well as a mutation operator to be applied at the location. The most common mutation operators are deletion, insertion, and replacement of fragments of code. A large population of candidate variants is produced through this process and validated against the test oracle to determine their fitness. The process is repeated until a suitable patch that passes all of the test cases is found.

In heuristic-based APR, the choice between the mutation operators in most approaches is random, causing the search to over-explore some operators that empirically lead to more failing software variants. As a result, more incorrect patches are produced and more resources are wasted which hinders the efficacy and efficiency of the technique. Smigielska et~al.~\cite{Smigielska2021} improve on this by implementing a uniform selection strategy where the probability for each operator to get selected is proportional to the size of its respective search space. Alternatively, Soto et~al.~\cite{Soto2018} use developer bug fix history to determine a fairer distribution of operators.

A variety of mutation operators is necessary for a diversified search space. However, if the application of the mutation operators throughout the search process is not optimised, this can create many incompatible variants which will delay or even prevent finding a correct patch. 
While previous work~\cite{Smigielska2021,Soto2018} does address this concern, the probabilities for each operator remain predetermined and fixed throughout the search. 
Therefore, the online feedback of the search is not used to further tune the probabilities for the selection of each operator. Moreover, the optimal probabilities for the different operators are likely different for every software. We aim to tune the operators accordingly so that the search behaves at its best for the software that is currently being processed. Given that evolutionary computation is often used in APR as the search strategy, we drew inspiration from there. We found that augmenting a reinforcement learning agent to optimise the selection process of the mutation operators has been implemented for evolutionary algorithms~\cite{DaCosta2008,Fialho2009,Fialho2010,Thierens2005,Murata1996,Eiben2007,J.E.Pettinger2002}, differential evolution~\cite{Sharma2019}, and evolutionary programming~\cite{Zhang2008}. To the best of our knowledge, this is yet to be applied for APR.%automated program repair. 

Our paper aims to tackle this gap in the literature by proposing a reinforcement learning based technique to inform the selection of mutation operators in heuristic-based APR to improve both the efficiency and efficacy of the process. 
We implement two credit assignment techniques: average and exponential recency-weighted average credit assignment and four reinforcement learning-based mutation operator selection strategies: probability matching, adaptive pursuit, epsilon-greedy multi-armed bandit, and upper confidence bound. 
We experiment with two types of rewards: raw fitness values and fitness values relative to the parent. The activation point of the technique during the search process is also configured to best fit the context. We analyse the efficacy and efficiency of the different techniques to find the best strategy for guiding the search process in heuristic-based APR. 

Our empirical results show that while the RL-guided mutation operator selection did generate more test-passing variants over the random one, it did not exhibit an improvement in the number of bugs patched.
% \todo{I think we need data here, I'll update once I read through results}
We hypothesize this is mainly due to the coarsness of the fitness function (Boolean value) which was not as effective in guiding the learning of the reinforcement agent.

%show that the RL-guided mutation operator selection approach are comparable to the traditional random selection strategy. Our contributions include:
% We find that epsilon-greedy mutation operator selection strategy is the most effective at generating patches for bugs while probability matching is the most efficient. Our results show that the RL-guided mutation operator selection approach improves on the standard random selection strategy by patching 9 additional bugs from a popular APR benchmark.
\begin{enumerate}
\item Novel reinforcement learning-based approach for mutation selection in heuristic-based program repair; %that outperforms the standard random selection techniques;
\item Extensible implementation of multiple selection techniques, credit assignment policies, reward calculations, and integration strategies in a state-of-the-art heuristic-based APR tool;\looseness=-1
\item Evaluation on 353 bugs from a popular APR benchmark.
% \item \todo{added this} Replication of results from the APR tool JarFly~\cite{Motwani2022}.
\end{enumerate}

\section{Background}
Next, we discuss the key ideas in heuristic-based automated program repair (APR) and reinforcement learning (RL) that are relevant as background for our proposed approach.\looseness=-1

\subsection{Heuristic-based Program Repair}
\label{sbpr}

Heuristic-based program repair 
navigates the search space of software variants in order to find software that fixes a given bug~\cite{surveyLeGoues}. The method is usually composed of three main stages. Given a buggy input program and its test oracle, the suspicious statements in the code are detected through the first fault localisation step. In the second stage, program variants are created by selecting mutation operators and applying them to the target locations to mutate them. Finally, the variants are all verified against the provided test oracle. A fitness function measures the viability of each variant based on the number of passing test cases and is used to guide the search strategy.

The two most commonly employed search strategies are local search~\cite{Qi2014} and genetic programming~\cite{Yuan2018}. Local search is a heuristic approach to optimisation problems in which small and random local modifications are iteratively applied to find better solutions~\cite{Hoos2004}. As for genetic programming, a population of candidate programs is created through evolving the source code using steps inspired by biological evolution~\cite{Koza1994}. 

%Our approach is fully tool-agnostic and can be augmented in any heuristic-based tool regardless of the search strategy that it implements. However, our review of APR tools in the literature (additional details in Section \ref{tool}) revealed that genetic programming is the more common strategy in state-of-the-art tools. Thus, we provide more details on this search strategy in this section.

Our review of APR tools in the literature (Section \ref{tool}) revealed that tools based on genetic programming are more common, thus we provide here more details on this search strategy.
 Traditionally, the genetic programming search strategy is compromised of 3 evolutionary steps: \textit{selection}, \textit{crossover}, and \textit{mutation}.
 In the \textit{selection} stage, the variants with the highest fitness in the population are selected.
 From there, the \textit{crossover} transformation is applied to the population of variants in pairs.
 The result of crossing over two variants is a new offspring variant.
 In the \textit{mutation} step, individual variants are mutated to create a second set of offspring variants. 

Most commonly, a program statement is mutated through a random choice between one of the following three options: deleting the statement entirely, inserting another statement chosen from elsewhere in the code after the current statement, or replacing the current statement with a different one.

More recent heuristic-based APR research improve upon this by using developer behaviour information during the search~\cite{Soto2019}, improving the fitness function~\cite{Le2016ml}, addressing issues of overfitting~\cite{Smith2015}, and proposing different mutation operators~\cite{Kim}.
However, the choice between the selection of the mutation options remains predominantly random with very few papers addressing improvements on this choice~\cite{Smigielska2021,Soto2018}. We propose the augmentation of reinforcement learning strategies to optimise the selection of mutation operators during the search process.\looseness=-1

\subsection{Reinforcement Learning}
\label{RL}

In reinforcement learning (RL)~\cite{Sutton2018}, sequential decision making is applied to problems where resources must be allocated between a finite number of conflicting choices. Given $N$ different action choices where only partial historical information is known about each, the idea is to predict the best possible series of actions in order to maximise the cumulative outcome. Ideally, over time as more choices are made, the overall gain is optimised. 

Balancing the exploration-exploitation tradeoff is at the core of RL techniques. \textit{Exploitation} of actions that the algorithm has sufficient historical information about allows for a reliable estimation of the expected rewards. However, \textit{exploration} is necessary in order to collect this information and optimise the selection of future actions. Choosing to exclusively exploit or explore actions will yield sub-optimal results. 

% \begin{sloppypar}
RL-based approaches have been used to improve the mutation operator selection stage in evolutionary computation in the past~\cite{Fialho2009,Maturana2009,DaCosta2008,Thierens2005}. In these works, the integration of such techniques is shown to outperform standard methods and optimise the selection of operators throughout the search. In this section, we detail the state-of-the-art RL algorithms that have been applied in this context as well as various variants for reward calculation and quality estimation from the literature.
% \end{sloppypar}

\subsubsection{Rewards}
\label{bg_Rewards}
Multiple variations have been suggested for the calculation of the reward values that the actions receive in the RL algorithm. This can be the raw fitness itself, but usually it is the fitness improvement in reference to another individual. The reference individual can either be the offspring's ancestor~\cite{relativeFitness2009}, the currently most fit individual in the population~\cite{Davis1989}, or even the median fitness of the population~\cite{Julstrum1995}. With this approach, an offspring that does not show an improvement is disregarded. In addition to just reflecting the fitness, the compass~\cite{Maturana2008} technique also takes into account the population's diversity.

\subsubsection{Estimating Action Qualities}
\label{calc_Q}

In this section, we discuss two techniques for estimating the qualities of actions~\cite{Sutton2018}. In the first technique, the estimated quality for an action \textit{A} at time step $t+1$ is the average of the rewards that the action has received until time step $t$ (Equation~\ref{averageQ}). This approach is appropriate for stationary environments where the distribution of the rewards is not altered over time. Alternatively, exponential recency-weighted average accounts for dynamic environments where the reward probabilities are non-stationary. With this approach, the estimated quality for an action \textit{A} at time step $t+1$ is calculated according to Equation~\ref{qualityTD} such that $Q_A(0) = 1$ and $R_A(t)$ and $Q_A(t)$ are the reward and predicted quality for action \textit{A} at time step $t$ respectively. This equation introduces hyper-parameter $\alpha$, which controls the learning rate.
\begin{align}
    \label{averageQ}
    Q_A(t+1) &= \frac{\sum_{i=1}^{i=t} R_A(i)}{t} \\
    \label{qualityTD}
    Q_A(t+1) &= Q_A(t) + \alpha[R_A(t) - Q_A(t)]
\end{align}

\subsubsection{Reinforcement learning algorithms} We present the state-of-the-art reinforcement learning techniques used for evolutionary strategies in the literature: probability matching, adaptive pursuit, and bandit-based algorithms.

\noindent
\textbf{Probability matching} (sometimes also referred to as Thompson Sampling or Posterior Sampling) is a simple technique that is widely observed in biological processes. With this approach, the probability of taking an action mirrors the probability of getting rewarded for it.
%Specific implementation details differ in the literature, however, in this paper we base the probability matching technique on the work by Thierens et~al.~\cite{Thierens2005} as their work applies to operator selection in genetic algorithms, often used in heuristic-based APR.
Specific implementation details differ in the literature; in this paper we follow Thierens et~al.~\cite{Thierens2005} as their work applies to operator selection in genetic algorithms, often used in heuristic-based APR.
To ensure that all actions continue to be explored throughout the search, they introduce a minimum probability value, $P\textsubscript{min}$. Given $P\textsubscript{min}$ and $N$ actions, the probability for an action \textit{A} at time \textit{t} with the estimated quality $Q_A(t)$ gets calculated according to Equation~\ref{PM_prob_formula}.
\begin{equation}
P_A(t) = P\textsubscript{min} + (1 - N * P\textsubscript{min}) \frac{Q_A(t)}{\sum_{i=1}^{i=N} Q\textsubscript{i}(t)}
\label{PM_prob_formula}
\end{equation}

\noindent
\textbf{Adaptive pursuit} (AP)~\cite{Thierens2005} aims to solve the probability matching slow convergence issue by introducing a ``winner takes all'' strategy. As with PM, the actions are selected according to their probabilities. However, the probability of an action \textit{A} at time step \textit{t+1} is calculated according to Equation~\ref{APEq}. The action \textit{M} is the action with highest quality value. This algorithm introduces a new hyper-parameter $\beta$ and a maximum probability value P\textsubscript{max}.
\begin{equation}
  P_A(t+1) = 
  \begin{cases}
    P_A(t) + \beta[P\textsubscript{max} - P_A(t)] & \text{if A = M}\\
    P_A(t) + \beta[P\textsubscript{min} - P_A(t)] & \text{ otherwise}
  \end{cases}
\label{APEq}
\end{equation}

\noindent
\textbf{Bandit algorithms} are the state-of-the-art approach to mutation operator selection in genetic algorithms.
In this paper, we experiment with two different variations: epsilon-greedy and upper confidence bound which have been shown successful in related work~\cite{Fialho2009,Maturana2009,DaCosta2008}.

The greedy bandit algorithm~\cite{Sutton2018} tends to favour exploitation of the best observed action thus far. Often as a result, the current best observed action keeps getting re-selected which prevents the algorithm from converging towards the best action. The epsilon-greedy~\cite{Sutton2018} improves on this by adding the option of exploration with a constant probability. With the epsilon-greedy algorithm, a random number between 0 and 1 is generated. If the generated number is less than $\varepsilon$, then an action is selected at random. Otherwise, the action with the highest quality (predicted reward) is greedily selected. Within bandit-based approaches the different action choices are referred to as \textit{arms}.

The upper confidence bound (UCB)~\cite{Auer2002} has been proven to optimise the cumulative gain and convergence rate. Given $N$ actions, Equation~\ref{UCB} presents the UCB action selection at time $t$ for action $A$ where $Q_A(t)$ is the action's estimated quality, $E$ is a constant balancing exploration-exploitation tradeoff, and $n\textsubscript{A}$ is the number of times that arm $A$ has been played. The chosen arm at time $t$ is the one that maximises the result of Formula~\ref{UCB}.\looseness=-1
\begin{equation}
    Q_A(t) + E * \frac{\sqrt{\log \sum_{j=1}^{j=N} n \textsubscript{j}}}{n\textsubscript{A}}
\label{UCB}
\end{equation}

\section{Approach}

In this work, we use reinforcement learning at the mutation operator selection stage of heuristic-based APR. 

\subsection{RL-Guided Mutation Operator Selection}
%Our approach uses reinforcement learning for the mutation operator selection process in heuristic-based APR. For each mutation operator, we track the fitness of the variants that it creates throughout the repair attempt. The fitness that we use to guide the reward update in our approach is the same fitness that guides the given APR search process. We use this to inform the probability of selection for each operator while balancing the exploration-exploitation tradeoff.\looseness=-1
Our approach uses reinforcement learning to augment the mutation operator selection process in heuristic-based APR. Every mutation operator is associated with a score used to guide the overarching APR search process. This score is updated every time this operator is selected using the fitness of the associated variant, thus improving following selections.

We implement the four most commonly used strategies in heuristic numerical optimisation problems: selection based on probability matching, adaptive pursuit, epsilon-greedy bandits, and upper confidence bound bandits. We chose these operator selection strategies, in particular, as these are simple, yet shown successful in the context of genetic algorithms~\cite{Thierens2005,DaCosta2008} --- a search strategy frequently used in heuristic-based APR.
We are also the first to apply these in this context, thus it is yet unclear which ones, or variants thereof, would be successful. For each of the above techniques, we experiment with an operator's reward as the raw fitness value or the fitness value relative to the parent and its credit being either based on exponential recency-weighted average or the average of the rewards (see Sections~\ref{bg_Rewards} and~\ref{calc_Q}). The variation in the credit assignment technique will inform whether the search process is stationary or dynamic.

This technique can be augmented into any heuristic-based APR tool that uses mutation operators. This is because we do not limit the number of actions (mutation operator options) to any specific value. We do, however, explore tuning the approach depending on the number of operators. We did this to investigate how to best apply the technique depending on the number of mutation operators the underlying tool implements.

\subsection{Credit Assignment}
\label{Credit Assignment}
When thinking about credit assignment in adaptive operator selection there are four main choices that must be made~\cite{Fialho2009}.

Firstly, \textbf{which type of reward will be used?} As this is the first paper to attempt reinforcement learning-based mutation operator selection, we experimented with the two most simple approaches: the raw fitness value of the individual and the relative fitness of the individual in reference to its direct parent. For an action \textit{A} at time step \textit{t}, we calculate the relative fitness $R_A'(t)$ according to Equation~\ref{relativeFitness}. This representation allows us to avoid negative numbers while still accounting for the exact level of improvement or deterioration in the offspring.\looseness=-1
\begin{equation}
  R_A'(t) = 
  \begin{cases}
    \frac{R_A(t)}{R_A(t-1)} & \text{if $R_A(t-1) \neq 0$,}\\
    R_A(t) & \text{ otherwise}
  \end{cases}
\label{relativeFitness}
\end{equation}

Secondly, \textbf{which operators to reward during the search?} Most commonly, only the operator that was used to create the offspring gets a reward. However, alternatives suggest rewarding the older ancestors of the offspring as well since they also had a part in its creation~\cite{Fialho2009}. Later work suggested that rewarding older individual's operators in that way is less effective~\cite{Barbosaancestor}. Therefore, we only award the direct operator that was applied to create the offspring.\looseness=-1

Thirdly, \textbf{how will the rewards accumulate throughout the search?} For each operator type, we can simply assign it its most recently received reward. We can also average all of the rewards that it has ever received. Since older rewards might be less relevant, Thierens et~al.~\cite{Thierens2005} assign a window with a fixed size \textit{N} where only the last N rewards contribute to the average. 
An improvement was made on this by Fialho et~al.~\cite{Fialho2009} to use an extreme-based credit assignment which is based on the assumption that infrequent large improvements in the fitness are more significant that frequent smaller ones~\cite{Whitacre2006}. We experiment with two simple strategies in this paper: using the average reward or the exponential recency-weighted average (see Section~\ref{calc_Q}). Using these strategies, we are able to calculate the estimated quality of each mutation operator. Experimenting with both of these strategies will help us better understand how stationary/dynamic the search process is in APR.\looseness=-1

Finally, \textbf{at which points in the search should the credit assignment occur?} An option would be to do so immediately after each operator is applied. This would mean that every time an operator is selected to create a variant, the variant's fitness would be evaluated and used to update the credit of the operator. Another approach would be to do this in batches, i.e., after a constant or variable number of variants. We experiment with both options.

\section{Research Questions}
In heuristic-based APR the variety of mutation operators increases the chances of patching bugs~\cite{Kim}. However, this is at the cost of a much larger search space, and a slower search as a result. We explore whether an RL-based mutation operator selection solution can better guide the search and mitigate this slowdown.\looseness=-1

\smallskip
\noindent \textbf{RQ1: Which credit assignment technique is suitable for mutation selection in heuristic-based APR?}
The choice of RL credit assignment strategy depends on whether the expected reward values can change over the course of the search. We compare the performance of two main stationary and dynamic methods to determine the environment type in heuristic-based APR.

\smallskip
\noindent \textbf{RQ2: Which mutation operator selection strategy is best in heuristic-based APR?}
RL has a long history of various selection strategies. We investigate the efficacy and efficiency of four different strategies that have been shown to be effective in similar contexts and compare with the standard operator selection technique in heuristic-based APR.

\smallskip
\noindent \textbf{RQ3: How does efficacy/efficiency of RL-based mutation selection in heuristic-based APR change with increase in the number of mutation operators?}
The addition of more fine-grained mutation operators that target specific bugs could improve efficacy but impede efficiency of traditional heuristic-based APR by creating a larger search space. We investigate whether this issue can be fixed by our RL-based approach.

\smallskip
\noindent \textbf{RQ4: To what extent does RL-based mutation operator selection improve the bug fixing ability of heuristic-based APR?}
Through online feedback of the search process, we hypothesise that the probability distribution of the mutation operators will be tuned to favour the more effective ones for the given buggy program, thus improving performance.

\section{Methodology}
\label{methodology}
Next we describe our methodology for answering RQs. 
First, we conduct a pre-study, where we fine-tune learning rate $\alpha$ (see Equation~\ref{qualityTD}) values for each of the four selection strategies. We use 10\% of the dataset for these preliminary experiments. While the optimal dataset size for parameter tuning remains an open problem, running these experiments on the full dataset would be too costly and would risk overfitting to the benchmark.

To answer RQ1, we run each RL algorithm with both of the credit assignment techniques on a variety of real-world bugs from an APR benchmark. For each operator selection strategy, we analyse its efficiency and efficacy when combined with the various credit assignment techniques. We then compare each of the optimal combinations of operator selection and credit assignment identified from RQ1 with the baseline, i.e., without using RL.
To answer RQ3, we explore the effect of changing the number of arms on the efficacy and efficiency rates of the proposed approach.

For the preliminary experiments and the first three research questions, we use the raw fitness value as the reward as it is the most simple strategy. We chose to assign credit to the operators at the end of each generation in batch at this stage to avoid too frequent updates that might overfit. Furthermore, most heuristic-based APR work relies on genetic programming which evolves populations of variants, already providing a natural division into batches.\looseness=-1 

For RQ4, we use the optimal settings of credit assignment and operator selection. In answering this question, we add experimentation for the reward types and integration strategy and compare our proposed approach with the baseline. 
We assess the quality of the patches by running the patched code on a second set of evaluation held-out test suites, which is a common strategy for patch evaluation in heuristic-based APR~\cite{Motwani2022,Soto2018,Soto2019} (further details on patch correctness evaluation in Section~\ref{experimentalSetUp}).
% \todo{Added ref to patch eval details as reviewer 3 misses that explaination}

We use the same measures for efficacy and efficiency to answer all of the research questions. Efficacy is measured by assessing the number of bugs for which a patch was generated, the frequency of such successful repair attempts for each bug, as well as patch quality, measured on held-out evaluation test suites. As for the efficiency measure we use the median and average numbers of individuals evaluated until a test-suite adequate patch is found.

\subsection{Tool}
\label{tool}

The field of APR has seen increasing growth in the last few years, with now tens of tools available\footnote{Refer to our tool comparison report for full details: \url{https://anonymous.4open.science/r/mutationLearner/reports/Tools.pdf}} to automate the task of bug fixing~\cite{program-repair.org:tools}.
After a thorough review of the heuristic-based APR literature, we found that JaRFly~\cite{Motwani2022} would be the best fit for our implementation. JaRFly is a novel open-source framework for search-based APR. 
It implements all three statement-level mutations that are used in GenProg~\cite{LeGoues2012bg} and TrpAutoRepair~\cite{Qi2013}: append, delete, and replace. It also implements 18 PAR templates~\cite{Kim} such as null checker and object initializer. JaRFly is well documented and very modular, making it an excellent candidate for extensibility.

When choosing a mutation operator to apply, JaRFly currently implements two options. The default is a uniform probability across all available mutations. They also allow for a probability distribution based on a probabilistic model. The model was created by mining open source repositories and analysing the frequency in which developers apply each of the available mutations~\cite{Soto2018}. We chose to use the uniform distribution approach as the baseline because the results that were reported show that operator selection informed by the probabilistic model generated a smaller number of patches than the uniform distribution~\cite{Soto2019}.

\textbf{JaRFly Modifications: } 
We implement a third reinforcement-based option in JaRFly that selects the mutation operator based on their current saved probabilities. These probabilities can be calculated according to any combination between the four operator selection strategies, the two credit assignment strategies, the two reward calculation techniques, and the two activation point variations. We implement each algorithm separately and depending on the mode with which the repair attempt was launched, the correct algorithm gets activated to set the probability values in the search process. We used the JaRFly fitness value as is in our implementation. The fitness value in JaRFly is calculated on the basis of the number of passing tests in the provided test suite.

Running the experiments revealed some bugs in the underlying code. We were able to locate five bugs that were causing uncaught exceptions in the execution of the code. These bugs are likely due to the changing version of external libraries that JaRFly depends on, e.g., in one of the cases this caused a change in the treatment of two-dimensional array identifiers. Patches for these bugs were added and can be found in our artefact. While JaRFly does implement 18 PAR templates, currently the most recent version has a bug that throws a NullPointerException within some scenarios of the application when 3 of the PAR templates generate variants, namely Parameter Replacer, Adder, and Remover. Therefore, we exclude these 3 PAR templates from all of our experiments.
It is important to note that this exclusion was across all of the experiments that we ran and thus does not invalidate the results we obtained.

\subsection{Benchmark}
\label{benchmark}
We considered various benchmarks for evaluation through the overview that~\cite{program-repair.org} provides. However, we chose Defects4J as it is currently the most comprehensive and popular dataset for evaluating Java APR tools~\cite{Martinez2016,Yuan2018} which allows us to be able to compare our results with the state-of-the-art. Additionally, Defects4J was used to evaluate the tool that we chose to implement our approach in, JaRFly~\cite{Motwani2022}, which allows for a direct comparison.

Our paper uses the most recent (2.0.0) version of Defects4J~\cite{defects4j} for evaluation. Defects4J is a collection of real-world Java bugs from open-source repositories. Each defect in the dataset includes the defective version of code, its developer-fixed version, as well as test oracles. There are two types of tests that accompany every Defects4J bug: developer written tests as well as the infrastructure for generating automated tests using EvoSuite~\cite{evosuite} or Randoop~\cite{randoop}.

To fairly evaluate our approach, we used the exact Defects4J bugs that were used for the baseline JaRFly program. At the time that the JaRFly framework was evaluated, version 1.1.0 was used which consisted of 395 bugs. Given that the Mockito project was excluded in the JaRFly paper, we did not consider it. Finally, we omitted the 4 bugs that have since been deprecated. We evaluated our approach on the 353 remaining bugs. The JaRFly paper reports to be able to successfully repair 49 out of these bugs using the GenProg algorithm and an additional 15 active bugs using PAR templates. The breakdown of the bugs is presented in Table~\ref{tab:defects4j}.

% \todo{added justification for why we didn't run full 835}
%We do not extend to the full dataset of 835 bugs from Defects4J. Doing so would have been an enormous waste of computational resources as it would require to run the baseline for comparison as well. However, an extension to more bugs is indeed possible as we do make all of our scripts and replication instructions public (refer to Section~\ref{dataAvailability}). We already conduct a very large-scale evaluation in this paper as we launch over $30,080$ repair attempts using the latest version of this state-of-the-art benchmark. On average, the runtime for each repair attempt is about one hour (more detail on this in Section~\ref{results}). Thus, experimentation in this paper can be estimated to have taken around $30,000$ hours! With a lot of parallelisation and using a node cluster, this translated into multiple months of actively running experiments.
We chose not to extend our experiments to the entire 835 bugs of the latest Defects4J dataset, as it would have more than doubled the already very consequent computational budget involved. Indeed, with over 30,080 repair attempts and an average running time of about an hour per attempt (see Section~\ref{results}) it amounts to about 3.5 years of continuous computation, or in our case multiple months of active cluster usage.\looseness=-1

\begin{table}
  % \caption{Breakdown of the Defects4J bugs that were used in the JaRFly study~\cite{Motwani2022} and the reported numbers for bugs patched using their tool with GenProg and PAR.} 
  \caption{Breakdown of the Defects4J bugs that were used in the JaRFly study~\cite{Motwani2022} and patched using their tool.}
   \label{tab:defects4j}
  \centering{
  \begin{tabular}{lccc}
    \toprule
    Project & Currently  & Patched   &  Patched \\
    & Active Bugs & with GenProg & with PAR \\
    \midrule
    JFreeChart & 26 & 6 & 7\\
    Closure Compiler & 131 & 15 & 20\\
    Apache Commons Lang & 64 & 9 & 12\\
    Apache Commons Math & 106 & 18 & 23\\
    Joda-Time & 26 & 1 & 2\\
    \midrule
    Total & 353 & 49 & 64\\
  \bottomrule
\end{tabular}}
\end{table}

\subsection{Experimental Set Up}
\label{experimentalSetUp}
\label{Parameter Tuning}

To answer our research questions, we ran the latest version of JaRFly.
We used the default uniform selection strategy to get a baseline, before enabling the implemented RL-approaches. \looseness=-1

\textbf{Search settings:} For each bug, 20 repair attempts were launched independently, to account for the heuristic nature of the underlying genetic algorithm. The search parameters were set to the same values as specified in the original paper of JaRFly~\cite{Motwani2022}. As such, each repair attempt was bound to 10 generations with a population size of 40. Since we were not able to run our experiments in the identical environment of JaRFly in terms of the hardware specifications, we eliminated the 4-hour timeout per repair attempt and ran all experiments to completion of the 10 generations instead. JaRFly provides a replication package~\cite{jarfly-rep} which includes the bugs that were successfully repaired as well as scripts for launching the tool to repair a specific bug. These scripts were used both for replicating their results and for conducting our experiments.

\textbf{Hyper-parameter settings}: The various reinforcement learning algorithms that we experiment with in this paper introduce hyper-parameters. Table~\ref{tab:hyperparams} depicts the different parameters that are associated with the different algorithms (see Section~\ref{RL} for more details on the parameters and formulas). The values of the parameters in the table are those used in the literature~\cite{Thierens2005,Fialho2009}.

As for the learning rate $\alpha$ (from the exponential recency-weighted average formula --- see Section~\ref{calc_Q}), we conducted preliminary experiments to tune its value for each experiment type on a subset of bugs from the Defects4J dataset. The subset includes 5 diverse bugs which constitute just over 10\% of the 49 bugs that the baseline in JaRFly successfully repaired. Each of these five bugs was the first in lexicographical order from those that the baseline in JaRFly correctly patched from each of the five projects (Table~\ref{tab:defects4j}). The five bugs we used in the experiment were: Chart 1, Closure~102, Lang 10, Math 18, Time 19. For each of the 4 operator selection strategies, we conducted four experiments using 4 learning rate values for a total of 16 preliminary experiments. The value that we found in the literature for the learning rate in this context is 0.8~\cite{Thierens2005}. We experimented further with the learning rates 0.2, 0.4, and 0.6. Each repair attempt was repeated 20 times, to account for the heuristic nature of JaRFly's genetic programming algorithm. 
Overall, in this preliminary study, we conducted 1600 repair runs.\looseness=-1

\begin{table}
  % \caption{
  % Hyper-parameter values for the probability matching (PM), adaptive pursuit (AP), epsilon-greedy, and upper confidence bound (UCB) algorithms; $N$: no. of mutations.} 
    \caption{
  Hyper-parameter values for the PM, AP, epsilon-greedy, and UCB algorithms; $N$: no. of mutations.} 
  \label{tab:hyperparams}
  \centering{
  \begin{tabular}{ccc}
    \toprule
    Name & Associated Algorithms & Value\\
    \midrule
    P\textsubscript{min} & PM, AP &  $\frac{1}{2N}$ \\
    P\textsubscript{max} & PM, AP &  $1 - (N - 1)*P\textsubscript{min}$ \\
    $\beta$ & AP & 0.8 \\
    $\varepsilon$ & epsilon-greedy & 0.2 \\
    E & UCB & 10 \\

  \bottomrule
\end{tabular}}
\end{table} 

\textbf{RQ1 and RQ2}: To answer the first 2 research questions, we activated our modified version of JaRFly with the 3 simple mutation operators from GenProg: insert, delete, replace (see Section \ref{sbpr}) as these are the most commonly used and would provide a foundation to build on for evaluation. The evaluation for these 2 RQs was on the 49 bugs that the original JaRFly paper reports to repair using the GenProg algorithm. As explained in Section~\ref{methodology}, at this stage we used the raw fitness value as the reward and assigned credits to the operators every generation.

\textbf{RQ3}: The experimentation for RQ3 was extended to include the 15 PAR templates that are implemented in JaRFly for a total of 18 mutation operators. Therefore, evaluation for RQ3 was conducted on all of the bugs that were repaired with either GenProg or PAR operators in JaRFly for a total of 64 bugs.
Experimentation for RQ1-3 is limited to a small number of bugs to avoid over-fitting and find the best RL strategy for the ultimate task of repair which we investigate in RQ4.

\textbf{RQ4}: Given that we update weights after each generation, we note that it might not be enough time to evaluate all 18 mutation operators and learn optimal rewards. Moreover, several of the 18 operators naturally belong into groups, e.g., some manipulate parameter values, while others are concerned with bound checks. Therefore in RQ4, we activate the RL algorithm in the search using 3 different sets of arms for further evaluation:

\begin{enumerate}
    \item 3 arms: 3 GenProg mutations
    \item 18 arms: 3 GenProg mutations + 15 PAR templates
    \item 7 arms: 3 GenProg mutations + 15 PAR templates aggregated into four groups.
\end{enumerate}
The third set of arms groups together PAR templates that pertain to bugs within a similar category. We divide the PAR templates into four groups: functions and expressions (FunRep, ExpRep, ExpAdd, ExpRem), bound and null checks (NullCheck, RangeCheck, SizeCheck, LBoundSet, UBoundSet, OffByOne), casting and initialisation (CastCheck, ObjInit, CasterMut, CasteeMut), and multi-line edits (SeqExch). The mutations are chosen and applied as normal in the search algorithm. However, each operator's credit gets assigned to its respective group. From there, an operator's probability for selection is what was assigned to the arm of the group that the operator belongs to. 

We experimented with two reward types: raw fitness and fitness relative to parent as well as two modes for assigning credit: credit assignment after every generation and after every mutation. 
We extend the evaluation to 353 bugs from the Defects4J benchmark.

% \todo{slight rewording of patch eval to highlight we indeed conducted this}
We evaluate the correctness of the patches produced by our approach according to the same methodology used in JaRFly~\cite{Motwani2022} for a fair comparison. The authors of JaRFly use two versions of the automated test generation tool EvoSuite~\cite{evosuite} to create held-out evaluation test suites. They created evaluation test suites for 71 defects from Defect4J that passed their criteria for statement coverage which are publicly available~\cite{jarfly-rep}. For each patch, if the defect had an evaluation test suite, we applied the patch to the defect and executed both sets of tests (v1.0.3 and v1.0.6) on the patched code. From there, we compared the quality scores of patches. The quality score for a patch is: $\frac{T\textsubscript{Pass}}{T\textsubscript{Total}}$ where T\textsubscript{Pass} is the number of tests that passed and T\textsubscript{Total} is the total number of test cases. 

%% We ran the experiment on a cluster of 1028 compute nodes. The experiments were assigned to a subset of these nodes with varying specifications. RAM size ranged from 16GB to 375GB and local SSD disk space ranged from 80GB to 780GB. All CPUs were from the Intel Xeon processors range from a variety of generations with the number of cores ranging from 4 to 48. This did not affect the results as we ran all repair attempts to completion of the 10 generations.
\textbf{Environment} Experiments were conducted on a cluster of 1028 compute nodes with varying specifications. RAM size ranged from 16GB to 375GB and local SSD disk space ranged from 80GB to 780GB. All CPUs were from the Intel Xeon processors range from a variety of generations with the number of cores ranging from 4 to 48. This did not affect the results as we ran all repair attempts to completion of the 10 generations.
%as we expect the test suites to be deterministic. 
%On average, the computation time for the experiments ranged from 45 minutes to 1.5 hours per repair attempt. However, for some it reached upwards of 12 hours. To run the preliminary experiments and answer all RQs, overall 30,080 repair attempts were run.
\section{Results}
\label{results}
Overall, a total of 30,080 independent repair attempts were conducted.
Computation time for each repair attempt ranged, on average, from 45 minutes to 1.5 hours, up to 12 hours in longest runs.

\begin{table}
\caption{
Preliminary experiment results. The table presents the experiment type, learning rate ($\alpha$), percentage of repair attempts that produced a test-suite adequate patch, number of bugs patched, average/median number of variants that were evaluated until a patch was found.}
\label{tab:tuning}
\centering{
  \begin{tabular}{ccccccccc}

    \toprule
    Repair  & $\alpha$  & Success  & Bugs  & Avg.  & Med.  \\
    Type  &  & Rate &  Patched (/5) & Variant & Variant\\
    \midrule
    Baseline &  - & 30\% & 4  & 177 & 171.5\\
    \midrule

    \multirow{4}{*}{PM}   &  0.2 & 33\% & 3 & 154 & 131\\
      &  0.4 & 27\% & 3 & 141 & 80 \\
      &  0.6 & 28\% & 3 & 178 & 93\\
      &\textbf{0.8} & \textbf{34\%} & \textbf{4}  & \textbf{136} & \textbf{138}\\
    \midrule

    \multirow{4}{*}{UCB}  &  0.2 & 36\% & 4 & 154 & 136\\
      &  0.4 & 34\% & 4 & 145 & 103.5\\
      &  0.6 & 36\% & 4 & 139 & 119.5\\
     & \textbf{0.8} & \textbf{38\%} & \textbf{4}  & \textbf{155} & \textbf{117}\\
    \midrule
    
    \multirow{4}{*}{AP}  &  \textbf{0.2} & \textbf{35\%} & \textbf{4} & \textbf{127} & \textbf{79}\\
      &  0.4 & 33\% & 3 & 130 & 78\\
      &  0.6 & 28\% & 3 & 141 & 77\\
      &  0.8 & 31\% & 3 & 142 & 79\\
    \midrule
    
      &  0.2 & 24\% & 4 & 136 & 123\\
      epsilon- &  \textbf{0.4} & \textbf{28\%} & \textbf{4} & \textbf{133} & \textbf{97.5}\\
     greedy &  0.6 & 30\% & 3 & 156 & 140\\
      &  0.8 & 18\% & 3 & 103 & 90\\
    
  \bottomrule
\end{tabular}}
\end{table}

\subsection{Preliminary Experiments}
\label{preResults}

The results of the parameter tuning preliminary experiments described in Section~\ref{Parameter Tuning} are depicted in Table~\ref{tab:tuning}. We used these experiments to determine the optimal learning rate value for credit assignment using the exponential recency-weighted average for each operator selection strategy independently.

All 16 of the experiments that we set up at this stage had both higher successful repair attempts than the baseline and lower medians for the repair variant number. For each operator selection strategy, we chose the learning rate value that maximised the number of successful repair attempts. Given this criteria, for the probability matching and UCB algorithms we found that the optimal $\alpha$ value is 0.8. As for adaptive pursuit, it is 0.2 and for epsilon-greedy it is 0.4. These are the experimental set ups that we proceeded with.% to the next stage.

\subsection{RQ1: Best Credit Assignment Technique}
\label{rq1}

The results of the experiments with average and exponential recency-weighted average credit assignment are presented in Table~\ref{tab:notdResults}. Interestingly, we can see that in all of the experiments the success rate of the repair attempts is higher with the average credit assignment than with the exponential recency-weighted average credit assignment. Moreover, we can see that in all of the experiments, except for AP, the number of unique bugs that were fixed is higher as well. Therefore, average credit assignment is best suited for PM, UCB, and the epsilon-greedy algorithms whereas for AP, the exponential recency-weighted average is better.

We investigated the discrepancy for the unique number of bugs patched in the adaptive pursuit results and found that there were 3 bugs that were successfully patched in combination with the exponential recency-weighted average credit assignment that were not patched in combination with the average credit assignment. However, for these bugs, only 1 or 2 repair attempts successfully generated patches. i.e., their patches are harder to find in the search space. AP with the average credit assignment might have failed to produce these patches due to limiting the number of repair attempts to 20 in our experiments. We can thus conclude that the search process in heuristic-based APR is stationary. 
\FrameSep3pt
\begin{framed}
\noindent \textbf{RQ1}: Average credit assignment is best suited for APR. Thus, the search process in heuristic-based APR is stationary.
\end{framed}

\subsection{RQ2: Best Mutation Operator Selection Technique}
\label{rq2}
To fairly compare with the baseline in answering RQ2, we tried to replicate the baseline results from the JaRFly paper by running the 49 bugs (Section~\ref{benchmark}) that the authors report to generate patches for using the GenProg setting. Our results are presented in Table~\ref{tab:baseResults}. When attempting to replicate the results, we were only able to generate a patch for 41 out of the 49 bugs. Since we used the exact search parameter settings and seeds that were used in JaRFly, we think that the discrepancy is likely due to the removal of the 3 PAR templates (as explained in Section~\ref{tool}) as well as mismatched versions (see details in Section~\ref{threats}), and heuristic nature of the search. The latest versions of JaRFly and Defects4J were used in our experiments. We observed variance even when the same seeds were used.\looseness=-1

To answer this question, we compared all of the RL algorithms with their optimal credit assignment technique that was identified in RQ1 (Table~\ref{tab:notdResults}). For PM, we can see that 41 unique bugs were patched which is the same as the baseline. However, the success rate of the overall repair attempts was 0.6\% lower. As for UCB, only 40 unique bugs were patched, but the overall success rate was 2.6\% higher. AP had the same number of unique bug patches as the baseline with a 0.3\% decrease in overall success rate. Finally, with the epsilon-greedy experiment 43 unique bugs were patched (2 more than the baseline) with a 4\% higher success rate for repair attempts. Given these results, it is evident that the epsilon-greedy algorithm with average credit assignment is the best reinforcement learning-based mutation selection strategy.

Since the bugs that were patched with each experiment varied, we had to look at the intersection of the 35 common bugs that were patched in all of the experiments including the baseline in order to conduct an efficiency analysis. The results of the overall comparison are presented in Table~\ref{tab:allCommonBugs}. We can see that PM with average credit assignment was the only experiment to achieve lower median and average numbers than the baseline for the variant numbers of successful patches. However, it is important to note that the success rate of the experiments varied. Therefore, while the epsilon-greedy experiment does have higher median and average values for the variant number of the patches, this might be due to the higher success rate which skewed the values.

\FrameSep3pt
\begin{framed}
\noindent \textbf{RQ2}: Epsilon-greedy is the most effective mutation operator selection strategy, while probability matching is the most efficient for heuristic-based APR.
\end{framed}

\begin{table}
\caption{
      RQ1 \& RQ2. Repair attempts with average and exponential recency-weighted average credit assignment.
 }
\label{tab:notdResults}
  \centering{%
  \begin{tabular}{ccccccccc}
    \toprule
    Repair  & $\alpha$  & Success  & Bugs  & Avg.  & Med.  \\
    Type  & &  Rate &  Patched (/49) & Variant & Variant\\

    \midrule

    \multicolumn{6}{c}{\textit{Average Credit Assignment}} \\

    \textbf{PM}  & - & 43.3\% & 41  & 99 & 52.5 \\
    \textbf{UCB}   & - & 46.5\% & 40 & 125 & 88\\
    AP & - & 45.1\% & 39 & 96 & 54\\
    \textbf{epsilon-greedy}  & - & 47.9\% & 43 & 112 & 81\\

    \cmidrule(lr){1-6}

    \multicolumn{6}{c}{\textit{Exponential Recency-Weighted Average Credit Assignment}} \\
    PM   & 0.8 & 43.2\% & 38  & 107 & 52\\
    UCB   & 0.8 & 44.1\% & 39 & 121 & 80\\
    \textbf{AP}   & 0.2 & 43.6\% & 41 & 101 & 62 \\
    epsilon-greedy   & 0.4 & 44.7\% & 43 & 109 & 73 \\
  \bottomrule
\end{tabular}}
\end{table}

\subsection{RQ3: Additional Mutation Operators}
Given that the epsilon-greedy algorithm with the average credit assignment was the best strategy in answering RQ2, we proceed with this technique for experimentation with additional operators.%\looseness=-1
 
We run the experiment using 15 PAR templates (recall Section~\ref{tool} where we detail that the remaining 3 templates were excluded in our study) and the 3 GenProg mutations for a total of 18 mutation operators. The experiment resulted in a success rate of 35.2\% and patches for 51 unique bugs. As for efficiency, the average variant number for variants that resulted in a patch was 131 and the median 87. When comparing with the JaRFly results with 18 arms in Table~\ref{tab:baseResults}, we can see that our approach did not improve the results of the standard selection strategy. Our approach with 18 arms was able to solve one bug fewer and the success rate for repair attempts was also 0.4\% lower. 

\FrameSep3pt
\begin{framed}
\noindent \textbf{RQ3}: With the addition of 15 arms, reinforcement learning-guided mutation operator selection does not patch more bugs than the standard operator selection strategy. \looseness=-1
\end{framed}

\subsection{RQ4: RL-Aided Mutation Operator Selection Performance}

The results of RQ3 evidence that a drastic increase in the number of mutation operators lowers the efficacy and efficiency of the approach. To control the number of arms in the RL algorithm, we decided to experiment with separating the PAR templates into groups (details in Section~\ref{experimentalSetUp}). We additionally experiment with a second type of reward which is based on the fitness of the variant in reference to its direct parent in the mutation step (recall Section~\ref{bg_Rewards}). Finally, we alter the rate of learning and experiment with two more aggressive approaches for the reinforcement learning. The first approach considers recalculating the mutation operator probabilities after every mutation instead of every generation. The second recalculates the probabilities every generation, but decreases the population within each one. Since our experiments set a bound of 40 on the population size and a bound of 10 on the number of generations, we flip these values and launch repair attempts bound to a population size of 10 over 40 generations.

\begin{table}
\caption{
    RQ2\&RQ3: Baseline Results: The table presents the results of the uniform operator selection baseline with the 3 GenProg mutations as well as with the additional 15 PAR templates.
 }
\label{tab:baseResults}
  \centering{
  \begin{tabular}{ccccccccc}
    \toprule
    Mutations  & Success  & Bugs  & Avg.   & Med. \\
          &  Rate &  Patched &  Variant & Variant\\

    \midrule
    GenProg & 43.9\% & 41/49 & 105 & 57\\
    \midrule
    GenProg  & \multirow{2}{*}{35.6\%} & \multirow{2}{*}{52/64} & \multirow{2}{*}{119} & \multirow{2}{*}{74} \\
    and PAR \\
    \bottomrule
\end{tabular}}
\end{table}

\begin{table}
\caption{
    RQ2. Results for the intersection of the 35 common bugs that were patched in all of the experiments.
 }
\label{tab:allCommonBugs}
\centering{
  \begin{tabular}{ccccccccc}
    \toprule
    Repair  & Credit  & Success & Avg.  & Med.  \\
    Type  &  Assignment & Rate & Variant & Variant\\
    \midrule
    Baseline & - & 58.6\% & 105 & 57.5\\
    \textbf{PM}  & A & 57.9\% & 97 & 50 \\
    UCB  & A & 64\% & 125 & 85.5\\
    AP & W & 58.7\% & 99 & 61\\
    epsilon-greedy & A & 63.9\% & 109 & 76\\
  \bottomrule
\end{tabular}}
\end{table}

\begin{table*}
\caption{
    RQ4. Results of mini-experiments on the sample of 5 bugs for 3, 18, and 7 arms.
    %% with 10 or 40 generations using raw and relative fitness values as the rewards.
 }
\label{tab:projExtensionTuning}
\centering{
  \begin{tabular}{ccccccccc}
    \toprule
    Arms & Generations & Activation Point & Reward Type & Success Rate & Bugs Patched & Avg. Variant  & Med. Variant \\
    \midrule

    \multirow{4}{*}{3} & 10 & Every Generation & Raw & 32\% & 4 & 176 & 123\\
    & 10 & Every Generation & Relative & 32\% & 4 & 218 & 152\\
    & 10 & Every Mutation & Raw & 0\% & 0 & - & - \\
    & 40 & Every Generation & Raw & 16\% & 3 & 158 & 120\\
    \cmidrule(lr){1-8}

    \multirow{4}{*}{18} & 10 & Every Generation & Raw & 31\% & 3 & 116 & 92\\
    & 10 & Every Generation & Relative & 31\% & 4 & 219 & 133\\
    & 10 & Every Mutation & Raw & 0\% & 0 & - & - \\
    & 40 & Every Generation & Raw & 15\% & 3 & 175 & 129\\
    \cmidrule(lr){1-8}

    \multirow{4}{*}{7} &\textbf{10}&\textbf{Every Generation} & \textbf{Raw} &\textbf{43\%} & \textbf{4} & \textbf{145} &\textbf{108.5}\\
    & \textbf{10} & \textbf{Every Generation} & \textbf{Relative} & \textbf{43\%} & \textbf{4} & \textbf{229} & \textbf{186}\\
    & 10 & Every Mutation & Raw & 0\% & 0 & - & - \\
    & 40 & Every Generation & Raw & 18\% & 3 & 178 & 101.5\\
    
    \bottomrule
\end{tabular}}
\end{table*}

\begin{table}
\caption{
    RQ4. Results for the epsilon-greedy operator selection algorithm with 7 arms (GenProg and grouped PAR Mutations) with raw/relative fitness run on the 64 bugs.
 }
\label{tab:finalResults}
\centering{
  \begin{tabular}{ccccccccc}
    \toprule
    Reward & Success  & Bugs  & Avg.  & Med.  \\
    Type & Rate  &  Patched  & Variant & Variant\\
        \midrule
    Raw  & 34.6\% & 48 & 125 & 82\\
    Relative & 38.1\% & 51 & 159 & 113\\
    \bottomrule
\end{tabular}}
\end{table} 

\begin{table}
\caption{
    RQ4. Comparison of quality values, and percentage of patches with 100\% quality between our approach and the quality results presented in the JaRFly paper~\cite{Motwani2022}.
 }
\label{tab:quality}
\centering{
  \begin{tabular}{lccccc}
    \toprule
        Repair & Min. & Mean & Median & Max. & 100\% Quality\\
        \midrule
        JaRFly GenProg & 64.8\% & 95.7\% & 98.4\% & 100\% & 24.3\%\\
        JaRFly PAR & 64.8\% & 96.1\% & 98.5\% & 100\% & 13.8\%\\
        \textbf{RL-based APR} & 63.5\% & 95.8\% & 98.5\% & 100\% & 34\% \\
    \bottomrule
\end{tabular}}
\end{table}

Table~\ref{tab:projExtensionTuning} presents the results of the mini-experiments that we ran to test the additional settings of grouped mutations, reward types, and aggressive learning. These experiments were on the same sample of five bugs that was used in the preliminary experiments for tuning the learning rate (recall Section~\ref{Parameter Tuning}). The results confirm that increasing the number of arms from 3 to 18 does hinder the performance of the approach. However, we can also see that keeping the 18 mutations but grouping them into 7 arms instead does greatly improve the results. The more aggressive learning achieved both through a smaller generation size or adjusting the operator probabilities after every mutation yielded worse results.

From here, we proceeded with additional experimentation on the two most successful experiments (emboldened in Table~\ref{tab:projExtensionTuning}).
Table~\ref{tab:finalResults} presents the results of these two experiments on the dataset of 64 bugs (that were reported as patched using GenProg and PAR in JaRFly). The experiment with the fitness relative to the parent as the reward achieved a 2.5\% increase in the success rate of the repair attempts and patched 3 additional bugs. Therefore, we decided to test it on the additional 289 bugs that the JaRFly reports to not be able to generate patches for with GenProg and PAR (recall Table~\ref{tab:defects4j}). Our approach produced 80 patches for 10 bugs from the set that JaRFly didn't patch.
We then reran those 289 bugs for the baseline and found 93 patches for 23 unique bugs. However, after manual investigation of the patches, we found that only 2 of the 93 patches in the baseline were correct and corresponded to 1 unique bug. None of the patches generated using the RL strategy were correct.

We follow the patch evaluation methodology of JaRFly~\cite{Motwani2022} to assess the quality of the patches generated by our approach. We assess the quality of 487 patches that correspond to the 51 defects that the epsilon-greedy operator selection algorithm with average credit assignment, relative fitness values as the reward and 7 arms generated on the benchmark of 64 bugs (recall Table~\ref{tab:finalResults}). Table~\ref{tab:quality} presents the results of the quality scores of the defects. We remove the 2 patches generated for the Time 19 bug from our analysis as they did not pass any test in the evaluation test suite. The results do not show improvement in the minimum, maximum, median, or average quality values. However, we can see that the percentage of patches that were evaluated to have 100\% quality is 34\% which is significantly higher than JaRFly.
% \todo{what do you think of answer to RQ4?}
\FrameSep3pt
\begin{framed}
\noindent \textbf{RQ4}: Reinforcement learning-aided mutation operator selection is comparable in terms of bus fixed to the baseline standard uniform selection approach, though we observe more test-passing variants generated.
\end{framed}

\section {Discussion}
We deduce that the environment in heuristic-based APR is stationary (recall Section \ref{rq1}), i.e., the optimal probabilities for the arms in the RL algorithms do not change over time. 
As a result, the longer the search progresses, the more likely that the probabilities will be tuned to these constant optimal values. \looseness=-1

% \todo{added paragraph on result explainability}
The probability matching algorithm had the lowest success rate in our experiments as it is the most simple approach (recall Section~\ref{RL}). Adaptive pursuit performed slightly better which can be explained by the improvement this algorithm accounts for in the convergence rate. The UCB algorithm improves the success rate even further since it has been proven to optimise the convergence rate regarding the cumulative gain~\cite{Auer2002}. However, to do so, this algorithm continues to tune the exploration-exploitation tradeoff as the search progresses. Given that the environment in heuristic-based APR is stationary, we hypothesize that this simply added noise to the learning process.
% In addition, UCB restricts the sampling to the best actions in exploration as opposed to the epsilon-greedy algorithm which chooses randomly between all of the actions. We hypothesize that the combination of these two factors given the stationary environment quickly
and made the UCB over-exploit.
% which resulted in the epsilon-greedy approach being the most fitting for the task.\looseness=-1

The results demonstrate (Section \ref{rq2}) that the most effective mutation operator selection strategy (epsilon-greedy) is not the same as the most efficient (PM) which stems from the exploration-exploitation tradeoff. Algorithms that favour exploitation are less effective at finding patches for difficult-to-patch defects since they do not explore the search space as extensively. However, their more exploitative nature will allow them to find the patches for easier-to-patch defects more quickly. 
% The hyper-parameter settings of the different algorithm also affects this tradeoff (recall Section~\ref{Parameter Tuning}).

Increasing the number of mutation operators from three to eighteen greatly increased the search space and effectively slowed our approach's learning rate. However, grouping mutations that target similar defect types together into a single arm representation within the RL algorithm showed an improvement as the power of the PAR fix templates was maintained while controlling the noise that the large number of arms added.
% Therefore, while our approach does not put a limit on the number of mutations operators, we do recommend the grouping method to control for the number of arms that the RL algorithm rewards. The success of the grouping method hints at the importance of defect type classification within APR. 

Results show that while the RL approach generated more test-passing variants, it did not significantly improve the number of bugs patched. We hypothesize that this is due to three different reasons. Firstly, since we limit the time budget of the experiments, the learning might have been too slow to begin taking effect. Thus, such an approach may only be effective in cases where the search is very long and the probabilities are learned quickly.
Secondly, the fitness function in this instance might be too coarse to effectively guide the learning.  
Finally, the type of edit may not be a sufficient source of information to steer the random edit generation toward the search sub-space containing the correct patch.

% \todo{weakened critisim of learning-based apr}
Out of all APR approaches to-date, heuristic-based techniques have been the most adopted in industrial settings~\cite{Kirbas2020,Janus,sapfix} (with \cite{sapfix} using multiple approaches). We pose this is largely due to these techniques not requiring a training stage, a pre-trained model, or constraint solvers. We recognise the potential in their continued scaling while accounting for the high success rates of end-to-end learning-based strategies and large language models~\cite{Sobania2023, Jiang2021,Zhu2021,Xia2022}.
% \todo{added cites from fse feedback}s
%We wanted to propose an approach that integrates a classical machine learning technique (RL) within heuristic-based automated program repair.
% Our results show that our technique surpasses all of the state-of-the-art tools in heuristic-based automated repair. Our approach generates patches for 61 bugs. Tools such as Arja~\cite{Yuan2018} and JaRFly~\cite{Motwani2022} generate patches for 59 bugs %\footnote{Evaluation does not include the Closure Compiler project from Defects4J} and 52 bugs\footnote{JaRFly reports to patch 64 distinct bugs with GenProg/PAR. We were only able to reproduce 52. ARJA evaluation does not include the Closure project.} respectively. While we do conduct this comparison given the results of our approach when implemented in JaRFly, our approach can be deployed in any heuristic-based tool regardless of programming language, repair granularity, or search strategy.  

\section{Related Work}
\label{related work}
% \todo{Reviewed new literature and added the new references}
%% Our literature review revealed four popular approaches for RL-based mutation selection. None of these, however, have been applied to improve mutation selection strategies for improvement of existing software. We are the first to do so for the problem of APR.
Our literature review revealed four popular approaches for RL-based mutation selection. None of these, however, have yet been applied to in the context of APR.
% \begin{sloppypar}

 \textbf{Mutation Operators:} Papers that investigate varying operator probabilities~\cite{A.Tuson1998,Murata1996} and tuning parameters generally~\cite{Aleti2016,Moazen2023} in the field of genetic algorithms have been around for decades~\cite{Stanczak1999,Julstrom1997}. These works generally focus on the adaptive operator selection based on a probability matching strategy that references the fitness improvements of the individuals~\cite{Vafaee2008a,Soria-Alcaraz2014}. Thierens et~al. propose an alternative to this through their adaptive pursuit strategy~\cite{Thierens2005}. Other papers focused on the mutation operator specifically both in the field of genetic algorithms~\cite{Hesser1991,Ali2013} and genetic programming~\cite{Anik2013,Hong2014}. The shortcomings of mutation operation, particularly in promoting diversity for GP~\cite{Jackson2011} are discussed in the literature as well. Glickman et~al.~\cite{Glickman2000} and Friedrich et~al.~\cite{Friedrich2018} address issues such as premature convergence and local optima escape.

Multiple studies alter operator probabilities using reinforcement learning~\cite{Yu2023RL, Yu2023Robust, Yu2024}. These studies address both credit assignment strategies as well as operator selection techniques, therefore they are particularly relevant to our paper~\cite{Awad2022,Li2023,Zhang2023}. More specifically, extensive research has been done on operator selection strategies that use multi-armed bandits~\cite{DaCosta2008,Fialho2009,Fialho2010,Thierens2005,Murata1996,Eiben2007,J.E.Pettinger2002,Maturana2009}. %However, none of these papers were aimed at APR. They were intended for the purpose of evolution strategies that mainly focus on numerical optimisation problems.
These paper mainly focus on augmenting evolution strategies in the context of numerical optimisation.

% \end{sloppypar}

In the field of APR specifically, we found some work by Le~Goues et~al.~\cite{LeGoues2012} on operator design choices. Smigielska et~al.~\cite{Smigielska2021} propose a uniform strategy when choosing operators. Enhancements were suggested using probabilistic models~\cite{Soto2018}, program context~\cite{Wen2018,Ullah2023}, and bug fix history~\cite{Le2016op}. Soto et~al.~\cite{Soto2019} further propose a set of techniques for enhancing the quality of patches. All of this work focuses on a fixed distribution of probabilities for operator selection. Our work takes this a step further by using reinforcement learning to modify this distribution during search.% on its online feedback.

 \textbf{Machine Learning for APR:} The utility of the machine learning component varies from predicting patch correctness~\cite{Schramm2017,Tian2023}, to predicting the type of fault~\cite{Valueian2022}, to predicting whether continuing the search process will result in a repair~\cite{Le2016ml}. Existing work guides the search process by learning from code patterns and features~\cite{Chen2022,Valueian2022}, bug reports~\cite{Liu2013}, program namespace~\cite{Parasaram2023}, or context and statistics~\cite{Jiang2019,Yu2023}. Ji et al.~\cite{Ji2022} applies program synthesis to the problem of automated program repair. Conner et~al.~\cite{Connor2022} use neural machine translation models that generate edit operations for patching bugs instead of translating from the buggy to fixed source code directly. 
 Additional work focuses on more specific types of repair, e.g., for conditional statements~\cite{Gopinath2016}, compilation errors~\cite{Ahmed2022}, API misuses~\cite{Wu2022}, or specific program languages~\cite{Lajko2022}.
 Finally, more recent work also include repair based on generative language models~\cite{Kang2023, Xia2023} as well as more interactive user-centric techniques~\cite{Liu2024, Geethal2023}.
 %% Our approach does not aim to target specific bug classes or languages and can be augmented into any heuristic-based APR technique to tune the probabilities of selection for mutation operators during search.

\section{Threats to Validity}
\label{threats}

\textit{External threats:} Defects4J is composed of real-world bugs and is a widely used benchmark in the literature. However, there always remains a threat of generalisability. Moreover, our methodology uses Java which may not generalise to other programming languages. Our choices of benchmark and language enable direct comparison with state-of-the-art approaches. We do not limit our approach to any specific benchmark, programming language, or tool therefore these threats can be mitigated through additional implementations.\looseness=-1

\textit{Internal threats:}
Our experiments were conducted on a cluster of computation nodes with varying specifications. We mitigated this threat by comparing the variant number of the successful patch instead of execution time to measure efficiency.\looseness=-1

\textit{Threats to construct validity:} The results of the JaRFly baseline were not reproduced despite using the same search parameters. Since our approach uses the most recent version of both JaRFly and Defects4J, the discrepancy is likely due to a version issue, the heuristic nature of search, and the omission of 3 of the PAR templates as explained in Section~\ref{tool}. Patch quality assessment remains an open question in the field. We mitigate the threat of potential overfitting by evaluating the patches on held-out evaluation test suites, and by manual evaluation of final test-passing variants.

\section{Conclusions and Future Work}

In this work, we introduce a reinforcement learning approach for mutation operator selection in heuristic-based automated program repair.
We conducted an extensive empirical evaluation, spanning four learning algorithms and two credit assignment techniques, and assessed the effect of reward types, number of mutation operators, and activation points on the performance.
Our findings reveal that this approach is comparable in terms of bugs fixed to the baseline standard uniform selection approach, despite RL-based selection generating more test-passing variants.
 
However, our analysis suggests several avenues for future investigation.
Firstly, the RL mechanism may not have sufficient time to significantly influence the probability of generating the correct edit within our budget-limited experiment.
Secondly, the fitness of variants may not serve as the most efficient reward source during learning.
Lastly, uncertainty remains regarding the effectiveness of edit types in guiding the search towards correct solutions.
 
In summary, our study provides valuable insights into RL-aided mutation operator selection whilst also underscoring the complexity inherent to automated program repair.
Future research efforts should strive to address these challenges, through innovative methodologies that balance computational efficiency with robustness in identifying correct patches.

\section{Data Availability}
\label{dataAvailability}
All source code, supplementary materials, and instructions needed to replicate our results are publicly available at: \url{https://anonymous.4open.science/r/mutationLearner}.

\bibliographystyle{ieeetr}
\bibliography{main}

\end{document}